# Calibrated absolute optical contrast for high-throughput characterization of horizontally aligned carbon nanotube arrays

Yue Li,[1,2] Ying Xie,[1] Jianping Wang,[3] Yang Xu,[2] Shurui Wang,[1,3] Yunbiao Zhao,[2] Liu Qian,[1] * Ziqiang Zhao,[1,2] * Jin Zhang[1,3] *

[1] School of Materials Science and Engineering, Peking University, Beijing, 100871, China.

[2] State Key Laboratory of Nuclear Physics and Technology, School of Physics, Peking University, Beijing 100871, P. R. China.

[3] Beijing Science and Engineering Center for Nanocarbons, Beijing National Laboratory for Molecular Sciences, College of Chemistry and Molecular Engineering, Peking University, Beijing, 100871, China.

E-mail: jinzhang@pku.edu.cn; zqzhao@pku.edu.cn; qianliu-cnc@pku.edu.cn

## Abstract

Horizontally aligned carbon nanotube (HACNT) arrays hold significant potential for various applications in nanoelectronics and material science. However, their high-throughput characterization remains challenging due to the lack of methods with both high efficiency and high accuracy. Here, we present a novel technique, Calibrated Absolute Optical Contrast (CAOC), achieved through the implementation of differential principles to filter out stray signals and high-resolution calibration to endow optical contrast with physical significance. CAOC offers major advantages over previous characterization techniques, providing consistent and reliable measurements of HACNT array density with high throughput and non-destructive assessment. To validate its utility, we demonstrate wafer-scale uniformity assessment by rapid density mapping. This technique not only facilitates the practical evaluation of HACNT arrays but also provides insights into balancing high throughput and high resolution in nanomaterial characterization.

**Introduction**

Horizontally aligned carbon nanotube (HACNT) arrays have garnered significant interest due to their unique physical and chemical properties, which are highly beneficial for applications in nanoelectronics, sensors, and composite materials[1,2]. Over the past few decades, the focus has shifted from laboratory research to practical applications, revealing challenges in achieving uniformity and stability at the wafer scale using traditional fabrication methods. To overcome these bottlenecks, it is crucial to develop characterization techniques that provide efficient and accurate feedback for experimental optimization. Traditional methods such as scanning electron microscopy (SEM), atomic force microscopy (AFM), and transmission electron microscopy (TEM) are precise but time-consuming and limited in throughput[3]. Therefore, there is a pressing need for a high-throughput, non-destructive technique to accurately characterize HACNT arrays.

Optical characterization offers high throughput and has been widely used in fields such as defect detection in electronics[4], biomedical imaging[5], and material science[6,7]. However, the nanoscale properties of carbon nanotubes (CNT) result in an extremely limited scattering cross-section in optical characterization[8,9], with contrast levels around $10^{-4}$. In our previous work, we used polarized optical microscopy to enhance the contrast of CNTs by two orders of magnitude, demonstrating the feasibility of optical methods for the evaluation of HACNT arrays[10]. Nevertheless, the exact correlation between optical contrast and density remained unclear. One major reason is the relative nature of contrast measurements, especially when transparent substrates like sapphire or quartz are used, where stray light from the substrate interferes with the CNT contrast. Additionally, our previous work used the absorption cross-section of carbon atoms and the average diameter of CNTs to derive the density from optical contrast without considering the complexity of the actual optical path, which led to deviations from the real density.

In this study, we introduce the Calibrated Absolute Optical Contrast (CAOC) method, a novel approach designed to address these challenges. First, we incorporated the angle between the HACNT arrays and the polarizer as a degree of freedom and obtained the optical contrast of CNTs. Using the differential principle, we extracted the cross terms between the CNTs and the substrate, naturally filtering out stray signals other than those from the CNTs. Due to its measurement consistency, we refer to this as absolute optical contrast (AOC). Furthermore, we calibrated the AOC using characterization techniques that can achieve single-tube resolution, establishing an absolute calibration function between density and AOC. This method significantly improves the characterization efficiency of HACNT arrays by approximately $10^3$ times, enabling high-throughput, wafer-scale sample analysis while maintaining the density measurement accuracy of high-resolution techniques.

**Principle and acquisition of AOC**

Figure 1a shows the experimental setup for characterizing HACNT arrays using polarized optical microscopy. The light source passes through the first polarizer ($P_1$) and illuminates the sample surface. Theoretically, the CNT-scattered electric field ($E_{NT}$) is polarized along the CNT direction due to a strong depolarization effect on light polarized perpendicularly to the CNT, while the substrate reflection ($E_R$) retains the polarization of $P_1$. Therefore, by setting $P_2$ perpendicular to $P_1$, substrate reflection is significantly reduced while preferentially retaining the CNT reflection, thus enhancing the contrast of the CNTs ($\Delta I/I_R$). As shown in Eq. 1[11], since $|E_{NT}|^2$ is extremely small, the contrast contribution mainly arises from the cross term. Hence, $P_1$ and $P_2$ need to be set at a small angle δ deviating from 90° to amplify the signal of the cross term. Experimentally, it was found that when δ is approximately ± 6°, the CNT contrast is maximized.

$$\frac{\Delta I}{I_R} = \frac{|E_R + E_{NT}|^2 - |E_R|^2}{|E_R|^2} = \frac{2|E_R| \cdot |E_{NT}|}{|E_R|^2} \qquad [1]$$

During the preparation of HACNT arrays, a-plane sapphire or ST-cut quartz substrates are typically employed for directional growth. Consequently, the impact of these transparent substrates, which exhibit polarization responses, must be considered during characterization. Reflections from the upper surface of the substrate maintain the polarization direction, whereas reflections from the lower surface undergo a polarization shift after traversing the substrate, thus escaping the cross-polarization setup and introducing substantial background signals[12]. To mitigate this issue, index-matching oil was utilized to eliminate interface reflections, and a wedge structure was employed to deflect the light, ensuring the CNT contrast remains at the order of 10$^{-2}$. This approach facilitates non-transfer characterization, which is essential for non-destructive sample testing. Figures 1b and 1c present optical microscope images of the same region under non-polarized and optimally polarized conditions, respectively, demonstrating a marked enhancement in CNT contrast with this optical configuration.

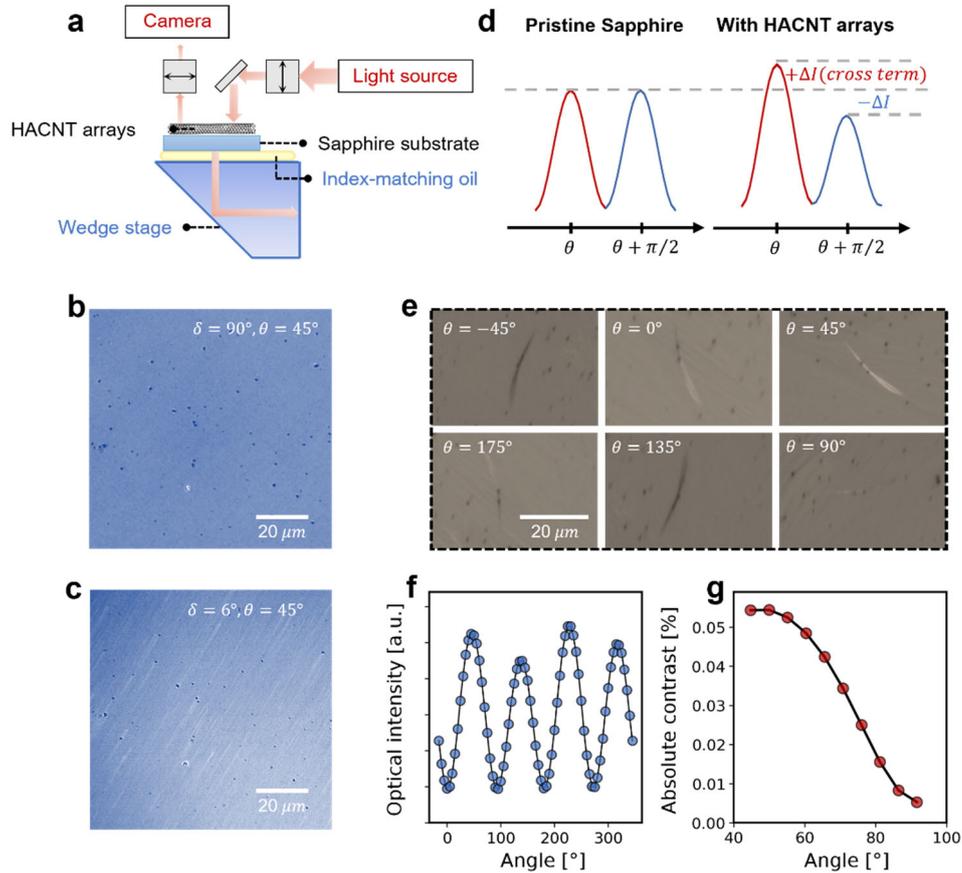

Figure 1. (a) Experimental setup for characterizing HACNT arrays using polarized optical microscopy. (b) Optical microscope image of HACNT arrays under non-polarized conditions. (c) Optical microscope image of HACNT arrays under optimally polarized conditions, showing enhanced CNT contrast. (d) Principle of absolute contrast measurement. (e) Dark-to-bright transition of a CNT bundle as the orientation angle θ changes. (f) Intensity-angle spectrum of HACNT arrays. (g) Absolute contrast results calculated using Eq. 3 based on the fitted data from (f).

The polarized optical setup enables the visualization of HACNT arrays using an optical microscope, however, accurate measurement of optical contrast is essential for quantitative characterization. Traditionally, contrast measurement relied on Eq. 2, which requires separately measuring both the total optical intensity ($I_T$) and the substrate optical intensity ($I_R$) to calculate the optical contrast ($\Delta I/I_R$). This method is suitable for single CNT or low-density arrays, but for high-density arrays, where CNTs fully cover the optical field of view, it necessitates the destructive process of scraping off nanotubes to create a substrate region (Fig. S1). Additionally, the complexity of transparent substrates often results in measurement errors, preventing accurate and consistent assessment of HACNT array optical contrast.

$$\frac{\Delta I}{I_R} = \frac{I_T - I_R}{I_R} \qquad [2]$$

To address these issues, we introduced the CNT orientation angle θ as a degree of freedom. By rotating the sample stage, an intensity-angle spectrum was obtained. Theoretically, the substrate signal exhibits a periodicity of $\pi/2$, while the cross-term signal between the substrate and CNTs displays a periodicity of $\pi$, changing sign every 90°. Fig. 1d illustrates this principle. The intensity-angle spectrum of pristine sapphire shows highly consistent $\pi/2$ periodic peaks, whereas samples containing HACNT arrays exhibit differences in peak heights, indicating the contribution of the cross-term, which is the primary source of CNT contrast (see Fig. S2 and Supplementary Note 1). Fig. 1e visually demonstrates this process, showing the dark-to-bright transition of a CNT bundle as θ changes. By processing the brightest and darkest points of the CNTs, other signals outside the CNTs can be naturally eliminated, yielding the absolute contrast of CNTs. Fig. 1f shows the actual measurement results of the HACNT array contrast spectrum, with peaks consistent with the theoretical predictions. Fig. 1g presents the contrast result calculated using Eq. 3 based on the fitted data from Fig. 1f. This method allows for the measurement of HACNT array contrast without separately measuring the background, significantly reducing measurement errors (Fig. S3).

$$\frac{\Delta I}{I_R} = \frac{I_T(\theta) - I_T(\theta + \pi/2)}{I_T(\theta) + I_T(\theta + \pi/2)} \qquad [3]$$

**Calibration workflow and establishment of AOC-density function**

AOC provides the theoretical basis for quantitatively characterizing HACNT arrays. In previous work, the absorption cross-section per carbon atom and the average diameter of the CNTs were used to correlate optical contrast with density[10]. However, this method is based on certain idealized assumptions, and the relationship between optical contrast and density necessitates further investigation. To address this, AOC was calibrated using high-resolution characterization techniques. Fig. 2 illustrates the calibration workflow, which primarily consists of two steps. First, we performed optical imaging at different angles, extracted the average brightness information from the images, and used Eq. 2 to obtain the AOC. Figs. 2a show optical images of HACNT arrays at two intensity peak positions, where the bright-to-dark transition is visible within the dashed boxes.

Subsequently, the same regions of the samples were characterized using AFM or SEM to count the number of CNTs. To ensure the feasibility of the calibration, the scales of the optical and high-resolution characterizations were unified. For instance, in AFM characterization, a scale of 2 μm was selected, and 25 consecutive images were taken to correspond to the information extracted from a 50 μm optical image. To enhance the efficiency and accuracy of CNT counting, a computer vision (CV) program was developed to identify and count the CNTs. Figure 2b visualizes the identification process, and Figure 2c presents the AFM results for a 50 μm region, with the horizontal axis representing the coordinates of the CNTs and the vertical axis representing the CNT count. The overall consistency of the slope indicates the uniformity of the sample. This figure can be infinitely magnified, as shown in the insets. Notably, using this method, the coordinates of each CNT can be recorded, similar to the infinite zoom of satellite imagery. This method also evaluates the uniformity of CNT pitch. Fig. S4 shows the pitch distribution of the CNTs obtained using "satellite imagery", which follows a log-normal distribution with a mean value and standard deviation of 20.52 nm and 12.22 nm, respectively. This precise characterization and analysis method may potentially facilitate the application of CNTs in carbon-based electronics.

By integrating AOC with "satellite imagery," the function between the exact density and optical contrast of HACNT arrays can be determined. As shown in Fig. 3a, the data points are derived from four different samples, with Sample 1 being a pristine sapphire. A strong linear relationship ($R^2 = 0.98$) is observed, consistent with previous reports. In contrast, using the method based on Eq. 2 to calculate optical contrast, the calibration (Fig. S5) shows that while the density and contrast of each sample are individually linearly correlated, there is no unified calibration relationship. This further underscores the consistency and reliability advantage of AOC.

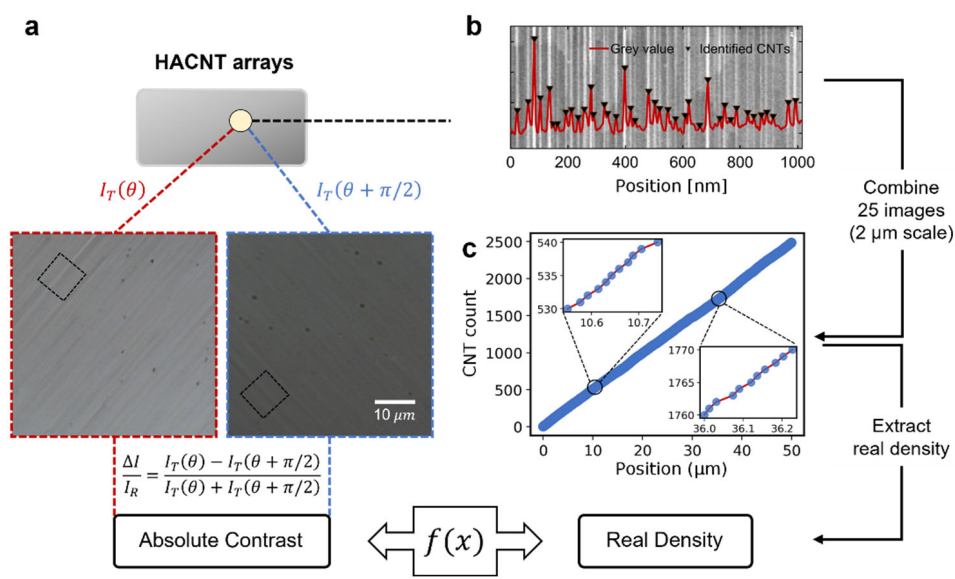

Figure 2. (a) Optical images of HACNT arrays at two intensity peak positions, showing the bright-to-dark transition within the dashed boxes. (b) Visualization of the CNT identification process using a computer vision (CV) program. (c) AFM results for a 50 μm region, with the horizontal axis representing the coordinates of the CNTs and the vertical axis representing the CNT count. Insets demonstrate the zoom-in of two areas.

The accuracy of the measurements was subsequently evaluated. Errors primarily stem from two sources: the accuracy of CNT identification by CV programs and the accuracy of the calibration curve. Figure 3b assesses the accuracy of CNT identification using 70 AFM images, with CNT quantity counted by both human scientists and CV programs (referred to as AI). The data points mostly fall on the y = x line, indicating the reliability of AI in density counting. Fifty points were selected to evaluate the accuracy of CAOC in density characterization. The horizontal axis represents the density results measured by AFM, while the vertical axis represents the density results measured by CAOC. As shown in Figs. 3c and S6, CAOC demonstrates overall good accuracy in density measurement, with larger errors in the low-density region (up to approximately 20%) and minimal errors in the high-density region. Given that the CNT signals are relatively weak even under polarized settings, the accuracy of this method is significantly influenced by density. Theoretically, the higher the sample density, the higher the measurement accuracy. Hereto, CAOC has been established as a novel method for high-throughput and accurate characterization of HACNT array density.

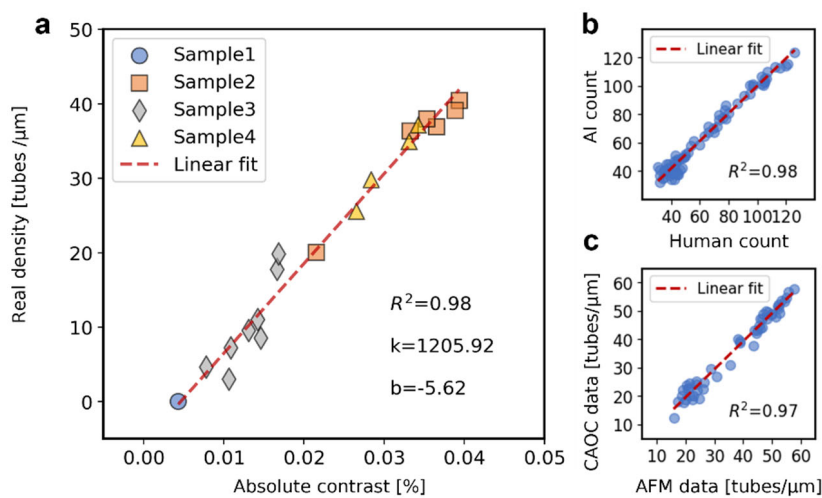

Figure 3. (a) The relationship between AOC and the real density of HACNT arrays. Sample 1 is a pristine sapphire. (b) Assessment of CNT identification accuracy, with CNT quantities counted by both human scientists and CV programs (referred to as AI). (c) Evaluation of CAOC accuracy in density characterization, with the horizontal axis representing density results measured by AFM, and the vertical axis representing density results measured by CAOC.

## Density mapping of large-scale HACNT arrays

A single AOC measurement requires taking 36 images, each with an integration time of 1 second. Considering the operator's handling time, one AOC point takes about 2 minutes. To further improve efficiency and support wafer-scale sample uniformity characterization, AOC was calibrated with optical intensity. For the same sample, during the same test (with consistent optical settings), $I_R$ is a constant value, so theoretically, optical intensity is proportional to AOC (Eq. 4 and SI). The measurement efficiency of optical intensity is 36 times that of AOC. Therefore, by selecting 5 points on a sample for calibration (approximately 5 minutes), the efficiency of subsequent density measurements can be significantly improved. Based on this, a standardized density characterization process was developed, as illustrated in Fig. 4a:

$$I_T \propto \frac{\Delta I}{I_R} \qquad [4]$$

1. High-throughput collection of optical intensity: Images of the HACNT array wafer were obtained using polarized optical microscopy with maximized CNT contrast. The average optical intensity was extracted from the central circular area (100 μm in diameter) of each image.

2. Calibration of the correlation between AOC and optical intensity: Five optical images were randomly selected, and the AOC in these regions was measured. The functional relationship between optical intensity and AOC of HACNT arrays was obtained through linear fitting.

3. Density calculation and analysis: Using the absolute calibration function and the

calibration function between AOC and optical intensity, all the optical intensities obtained in the first step were translated into HACNT array densities. This allowed us to statistically analyze the density uniformity of the HACNT-array wafer.

Fig. 4b illustrates the relationship between optical intensity, AOC, and actual density. Although the two calibration steps—from optical intensity to AOC and then to real density—can amplify experimental errors, this method significantly enhances efficiency, enabling practical assessment of wafer-scale density uniformity. Direct calibration between optical intensity and real density could further improve accuracy, but the calibration with high-resolution techniques required several hours.

Using this method, we characterized a 1 cm × 1 cm sample, uniformly selecting 144 (12×12) points across the sample and randomly choosing five points to measure AOC. The calibration curve is shown in Fig. 4c. Fig. 4d displays the density distribution of the entire sample, with an average density of approximately 40.40 tubes/μm and a standard deviation of about 7.95 tubes/μm. Fig. 4e is a mapping of the sample, illustrating that this method allows for rapid density partitioning, enabling the categorization of HACNT arrays for different applications. The entire characterization process took about 30 minutes, covering 1.44% of the sample area. In contrast, AFM characterization in the same time would cover only about $10^{-5}$ of the area, lacking statistical significance.

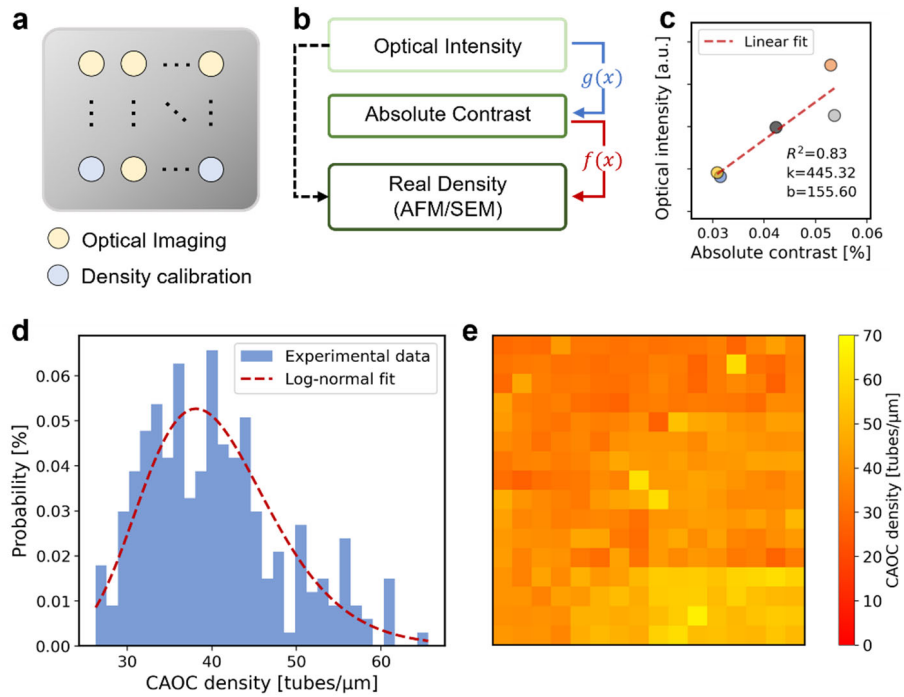

Figure 4. (a) Schematic of the standardized density characterization process for HACNT arrays using optical imaging. (b) Relationship between optical intensity, AOC, and real density of HACNT arrays. (c) Calibration curve of the relationship between optical intensity and AOC. (d) Density distribution of a 1 cm × 1 cm HACNT array sample. (e) Density mapping of the 1 cm × 1 cm HACNT array sample.

**Conclusions**

In this study, we developed a CAOC method for high-throughput and accurate characterization of HACNT arrays. We proposed a novel approach for calculating the optical contrast of CNTs by utilizing the differential principle to naturally filter out stray signals, thereby obtaining experimentally consistent AOC. By integrating AOC with high-resolution techniques, we established a calibration function between AOC and real density, demonstrating a strong linear relationship consistent with theoretical predictions. The CAOC method enables practical assessment of wafer-scale density uniformity. Our standardized density characterization process allows for rapid and reliable density mapping of large-scale HACNT arrays. The ability to perform high-resolution, non-destructive, and high-throughput density characterization represents a substantial advancement in the field of CNT, potentially accelerating the development of HACNT arrays and their applications.

However, the CAOC method has some limitations. For instance, it tends to have larger errors with low-density HACNT arrays, where optical contrast variations are minimal. This issue can be mitigated by using higher precision cameras. Additionally, AOC is not only related to density but also to the average diameter of CNTs[10]. Different growth methods may result in varying diameter distributions, necessitating extra calibration of the AOC-density function. Future work will focus on refining the technique to extend its applicability across a broader range of CNT characterization. Overall, CAOC provides a powerful tool for the rapid and accurate characterization of HACNT arrays, and it also shows potential for application in other one-dimensional materials. The principles of differential measurement and calibration can solve the challenge of balancing high throughput with high resolution, which is insightful for the development of nanomaterial characterization techniques, facilitating further research and development in nanotechnology.

## Methods

**Synthesis of HACNT arrays**

HACNT arrays were grown on a-plane sapphire substrates. The catalyst was prepared by ion implantation, followed by carbon nanotube growth using a home-made vertical spraying chemical vapor deposition (VSCVD) system, operating at 800-900 °C. Growth began with a flow of 300 standard cubic centimeters per minute (sccm) argon and 300 sccm hydrogen for 1-5 minutes to reduce Fe ions and form Fe nanoparticles in-situ. Subsequently, 10-20 sccm argon was introduced through an ethanol bubbler as the carbon feedstock. Throughout the process, argon flow was maintained at 300 sccm, and hydrogen flow varied from 300 to 500 sccm. The pressure remained at 70 Kpa, and the spraying distance was set at 10 mm.

**General characterization**

SEM images were obtained using a Hitachi S4800 SEM operated at 1.0 kV and 10 kV. Raman spectra of HACNT arrays with line mapping, conducted with a scan step of 5 μm and a laser beam spot of ~1 μm, were collected using a Jovin Yvon-Horiba LabRam system with 532 nm excitation. AFM images were obtained using a Dimension Icon microscope (Bruker).

**Optical characterization**

As shown in Figure 1a, an optical microscope (Nikon Eclipse LV100N POL) equipped with a 10× objective (N.A.=0.3, air), 40× objective (N.A.=0.6, air), 100× objective (N.A.=0.9, air), and a 50× objective (N.A.=0.5, 22 mm long working distance), two polarizers (Nikon LV-PO), and a color camera (Nikon DS-Ri2) were used to capture the polarized optical images. One polarizer (P1) was located in the incident beam, and the second polarizer (P2) was placed in the reflective beam, allowing free rotation to change the analyzing direction with a rotation precision of

0.1°. A halogen lamp was used as the light source (Nikon LV-LH50PC). The aperture iris diaphragm and the field iris diaphragm enabled convenient control of the incident beam, with the field iris diaphragm used to eliminate excess light. For birefringent substrates, refractive index matching was performed to suppress the depolarized light from the back surface. The typical image exposure time was 1.0 s, with an analog gain of 14.0x.

**Optical image processing**

All characterized images were processed using custom-developed Python or MATLAB programs. CNT identification was primarily based on image enhancement and peak-finding algorithms. Optical image processing involved averaging the RGB intensity within masked regions. The specific code will be released on GitHub after peer review.

### Data availability

The data that support the findings of this study are available from the corresponding authors upon reasonable request.

### Acknowledgements

This work was financially supported by the Ministry of Science and Technology of China (2022YFA1203302, 2022YFA1203304 and 2018YFA0703502), the National Natural Science Foundation of China (Grant Nos. 52021006, 52102032, 52272033), the Strategic Priority Research Program of CAS (XDB36030100), the Beijing National Laboratory for Molecular Sciences (BNLMS-CXTD-202001) and the Shenzhen Science and Technology Innovation Commission (KQTD20221101115627004).

### Author contributions

Y.L. and Y.X. prepared the samples and performed the experiments of synthesizing and characterizing HACNT arrays, assisted by J.W., Y.X., and S.W. Y.L. and L.Q. wrote and revised the manuscript with input from all authors. J.Z., L.Q. and Z.Z. supervised the overall projects. All authors contributed to the discussion and completion of this manuscript.

## Competing interests

The authors declare no competing interests.

# Supplementary Information

# Calibrated absolute optical contrast for high-throughput characterization of horizontally aligned carbon nanotube arrays

Yue Li,[1, 2] Ying Xie,[1] Jianping Wang,[3] Yang Xu,[2] Shurui Wang,[1, 3] Yunbiao Zhao,[2] Liu Qian,[1] * Ziqiang Zhao,[1, 2] * Jin Zhang[1, 3] *

[1] School of Materials Science and Engineering, Peking University, Beijing, 100871, China.

[2] State Key Laboratory of Nuclear Physics and Technology, School of Physics, Peking University, Beijing 100871, P. R. China.

[3] Beijing Science and Engineering Center for Nanocarbons, Beijing National Laboratory for Molecular Sciences, College of Chemistry and Molecular Engineering, Peking University, Beijing, 100871, China.

E-mail: jinzhang@pku.edu.cn; zqzhao@pku.edu.cn; qianliu-cnc@pku.edu.cn

**The supplementary information includes:**

Supplementary Figure 1-6

Supplementary Note 1-2

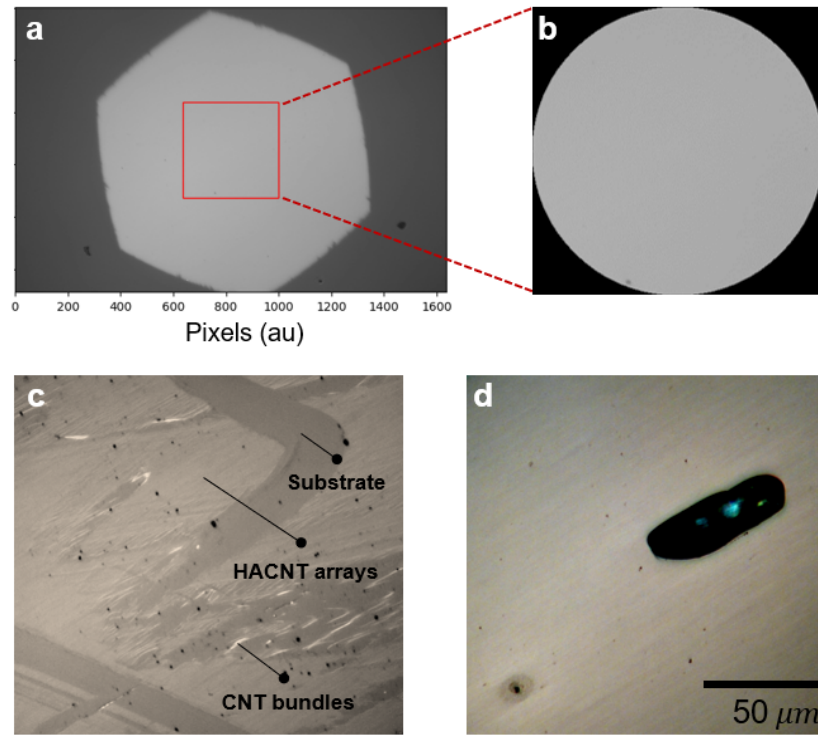

**Supplementary Fig. 1 |** Optical characterization details of HACNT arrays. (a) Unprocessed optical image with the horizontal axis representing pixel points (0.54 μm/pixel). (b) For AOC calculation, a mask is used to extract the circular region at the center of the optical image to obtain the intensity, ensuring a consistent characterization region during θ rotation. (c) Substrate region of HACNT arrays after being scratched with tweezers, which causes damage and generates CNT bundles. (d) Optimal polarized optical imaging of HACNT arrays, showing a selected region containing particles and contaminants, which appear black under polarized settings.

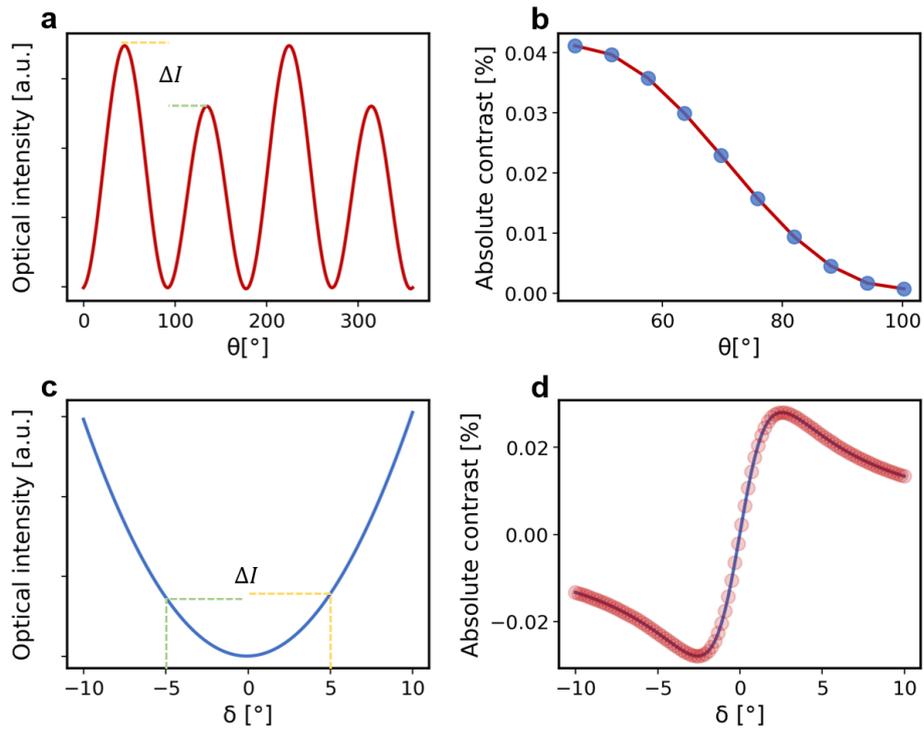

**Supplementary Fig. 2 | Numerical simulation of AOC.** (a) Intensity-angle spectra of HACNT arrays at different θ obtained through numerical simulation. (b) AOC at different θ derived from the data in (a) using the differential method. (c) Intensity-angle spectra of HACNT arrays at different δ obtained through numerical simulation. (d) AOC at different δ derived from the data in (c) using the differential method. The formulas used for the numerical simulation are detailed in Supplementary Note 2.

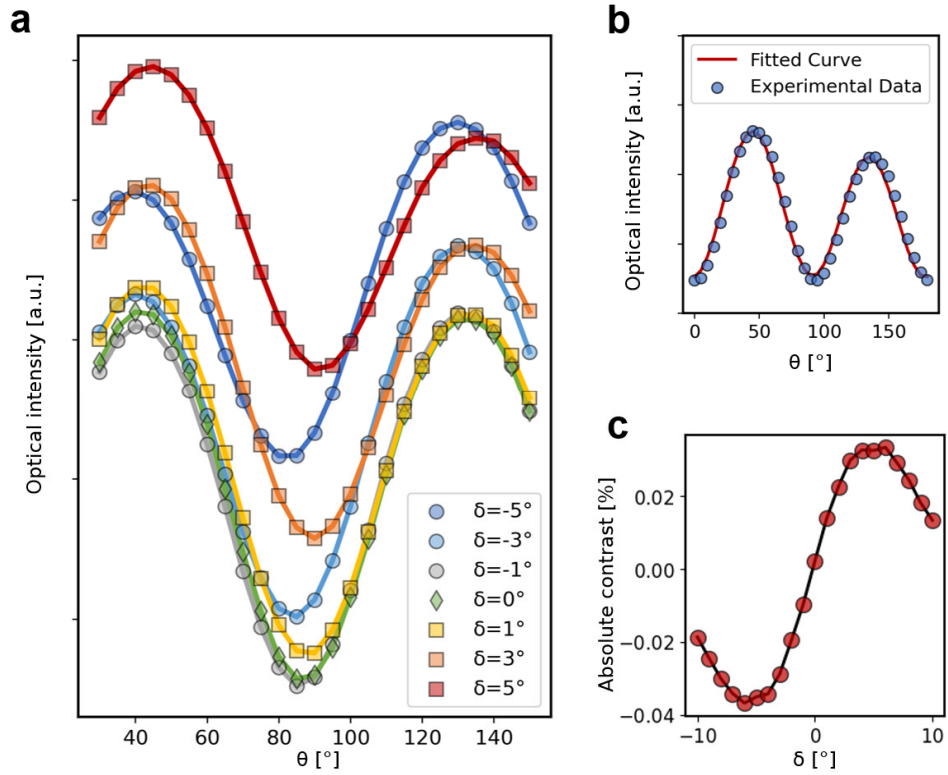

**Supplementary Fig. 3 | Measurement of AOC.** (a) Intensity-angle spectra of HACNT arrays at different δ angles. (b) Measured intensity results and theoretical fitting results, with the fitting formula provided in Supplementary Note 2. (c) Two peaks of each spectrum in (a) were identified through the fitting results, and AOC at different δ angles was calculated using Eq. 3, showing good agreement with theoretical and literature results.

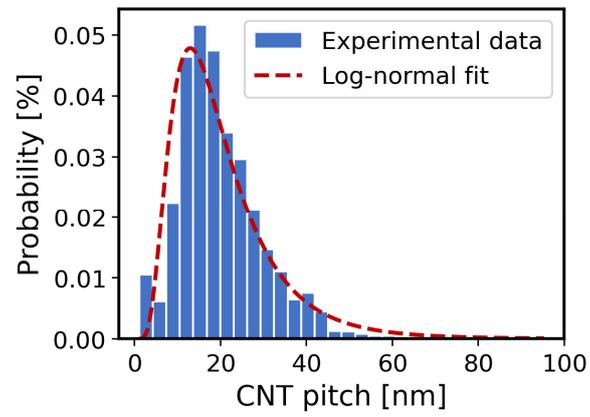

**Supplementary Fig. 4 |** Pitch distribution of the CNTs obtained using "satellite imagery".

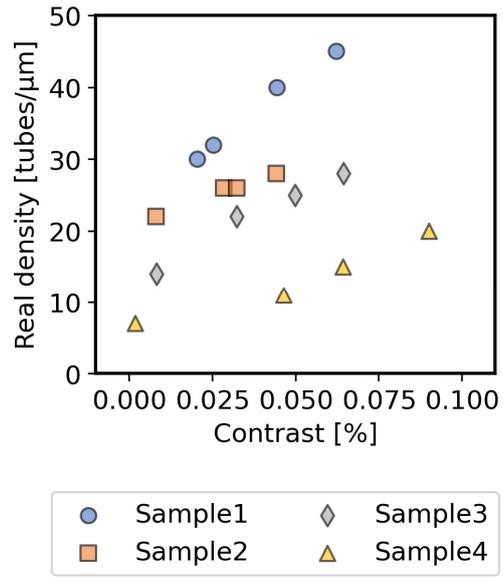

**Supplementary Fig. 5 |** The relationship between optical contrast measured by traditional method and the real density of HACNT arrays.

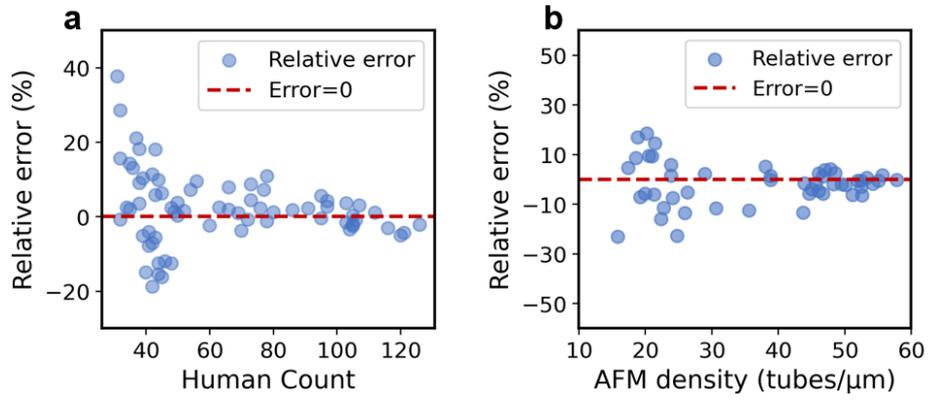

**Supplementary Fig. 6 | Evaluation of the accuracy of CNT identification and density characterization.** (a) Relative error in CNT density identification using the CV program. (b) Relative error in density characterization using CAOC.

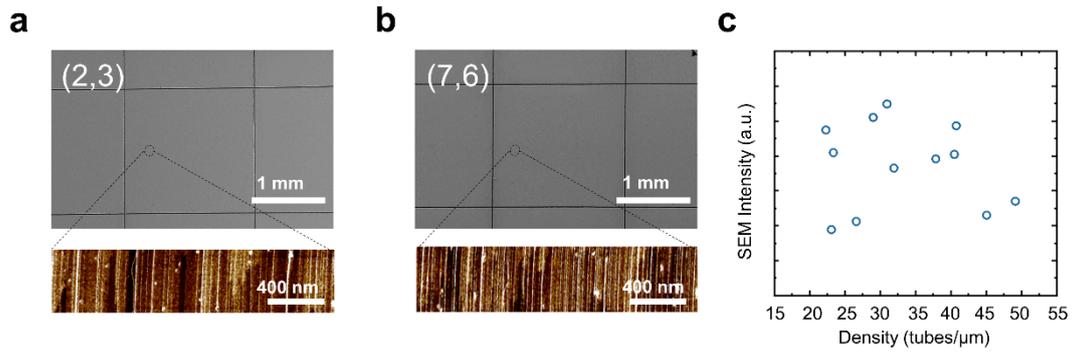

**Supplementary Fig. 7** | SEM characterization of HACNT arrays. (a-b) SEM images of regions two random selected regions on a HACNT array wafer, with actual density determined by AFM characterization. (2, 3) and (7, 6) is the coordinates of the selected regions. (c) Quantitative assessment of density characterization using SEM.

**Supplementary Note 1. Theoretical derivation and numerical simulation of AOC**

In our polarization setup, two nearly vertically crossed polarizers were placed in the incoming and outgoing light paths (referred to as P1 and P2). P1 and P2 have a small angle δ deviating from 90 degrees, and the angle between the CNTs and P1 is θ. For HACNT arrays on transparent substrates, the intensity ($I_T$) on the detector can be expressed as[1–4]:

$$I_T \cong |E_{NT} + E_R|^2$$

Further derivation yields:

$$I_T \cong E_0^2 \left[ \sin^2 \delta \cdot \sin^2 \theta + g \cdot \left(\frac{\alpha}{2}\right) \cdot \sin \delta \cdot \cos \theta \cdot \sin(\theta - \delta) + f\left(g, \frac{\alpha}{2}, \sin \delta, \theta, \delta\right) \right]$$

Where $E_0$ is the electrical field of the incident light after P1, $\alpha$ is the absorption cross-section of the CNT, and $g$ is a factor considering the proportion of the substrate signal to the CNT signal, which can be understood as the coverage of CNTs on the substrate, related to the density. The terms in brackets consist of three parts: the first term contributed by the substrate, the second term is the cross-term between the CNTs and the substrate, and the third term is higher-order infinitesimal term. Using this formula for numerical simulation, we can obtain the relationship between the optical intensity of HACNT arrays and the variations of θ or δ (Figs. S2a and c), which matches the experimental results.

For HACNT arrays on transparent substrates, the optical contrast of CNTs mainly results from the interference between the CNT-scattered electric field $E_{NT}$ and the substrate-reflected electric field $E_R$. Therefore, to obtain the CNT contrast, we need to extract the second term. In previous methods, this was achieved by separately measuring the substrate contribution to obtain the cross-term optical intensity. In fact, since the substrate signal exhibits a periodicity of π/2 (term 1), while the cross-term signal between the substrate and CNTs displays a periodicity of π, changing sign every 90° (term 2), we can use the differential principle to extract the second term without separately measuring the substrate signal, which is one of the main sources of error. This can be expressed by the following formula:

$$\frac{\Delta I}{I_R} = \frac{I_T - I_R}{I_R} = \frac{I_T(\theta) - I_T(\theta + \pi/2)}{I_T(\theta) + I_T(\theta + \pi/2)} = \frac{I_T(\delta) - I_T(-\delta)}{I_T(\delta) + I_T(-\delta)}$$

Using this method, we can derive the optical contrast, or absolute optical contrast (AOC), from the intensity-angle spectrum. The results of the numerical simulation are shown in Figs. S2b and d. Experimentally, we found that the differential process based on θ has better accuracy. Therefore, the CAOC method was developed based on adjusting θ.

**Supplementary Note 2. Trade-offs in choosing a high-throughput characterization method to determine the density of HACNT arrays**

Normally, we used the local highest density to depict the HACNT arrays. AFM and TEM were adopted for its high precision within a comparable characterization scope (tubes per micrometer)[5,6]. However, the low throughput limits its effectiveness in assessing the density distribution of HACNT arrays in wafer scale. Consequently, we employed SEM for the preliminary characterization of the wafer uniformity. As shown in Supplementary Fig. 7, we randomly selected and characterized two regions of the sample, (2,3) and (7,6). SEM characterization demonstrated good uniformity of the sample. However, as shown in the insets of Supplementary Figs. 7a-b, zoom-in characterization by AFM revealed distinct differences in the density of HACNT arrays between these two regions, with densities of approximately 20 tubes/μm and 40 tubes/μm, respectively. Furthermore, we extracted the gray values from several randomly selected regions in SEM images and compared them to the density of HACNT arrays in the same regions measured by AFM. These two sets of data were plotted in Supplementary Fig. 7c. Unfortunately, no correlation was observed.

Our results demonstrate a good linear relationship between optical characterization and density. So, why did SEM fail while optical methods succeeded? We speculated that there are three main reasons for the failure of SEM: (1) The imaging mechanism of CNTs in SEM was quite complicated, leading to an unclear relationship with multiple factors coupled between the density of HACNT arrays and the intensity in SEM images. With low-voltage SEM (1 keV in our experiments),

CNTs on insulating substrates showed bright contrast and their diameters were broadened to tens of nanometers. Several mechanisms have been proposed to explain this phenomenon, such as the electron-beam-induced current (EBIC) and the voltage contrast for local potential difference[7]. No matter what, the proposed mechanisms all involve multiple physical processes, such as the charging of insulating substrates, the charge transfer between CNTs and substrates, and the secondary electron emission which is very sensitive to the surface potential. The above-mentioned complicated and unclear physical processes lead to a complicated relationship between the density of HACNT arrays and the secondary electron intensity in SEM, which could be very difficult to unravel. (2) The SEM contrast contribution of an individual CNT is sensitive to its band gap[8]. Metallic and semiconducting CNTs on insulating substrates were found to show different contrast in one SEM image, which is mainly related to the potential of the charge transfer from CNT to substrate. This mechanism reveals that CNTs with different structures contributed differently to the SEM contrast. Only if we knew the exact band-gap distribution of the as-grown HACNT arrays could we roughly unravel the density of HACNT arrays from the SEM intensity. However, it is still less likely to realize in the current field of synthesis. (3) The background intensity of an SEM image is hard to keep stable as the scanning prolonged. For HACNT arrays grown on insulating substrates, the surface potential changed rapidly as the shoot of electrons went on, leading to an unstable SEM intensity. Thus, it is meaningless to make quantitative comparisons between different SEM images. Only if we kept the gray value of the background (insulating substrates in this case) exact identical, which was almost impossible for the SEM imaging.

Compared to SEM imaging, the physical processes involved in optical imaging of CNTs are significantly simpler, making it easier to establish a relationship between optical contrast and HACNT arrays density. Furthermore, the interaction between light and HACNT arrays will not interfere with subsequent detection. The optical contrast is considerably more stable over time compared to SEM imaging. Considering its high throughput, convenient operation, and stability, we believed that optical microscopy is the most promising technique for large-scale characterization of HACNT arrays density.